\documentclass[aps,prd,onecolumn,groupedaddress,showpacs,nofootinbib,amssymb]{revtex4}
\usepackage[dvips]{graphicx}
\usepackage{amssymb}
\usepackage{amsmath}
\usepackage{graphicx,,color}
\usepackage{amsfonts}
\usepackage{bm}
\usepackage{cancel}
\usepackage{comment}
\usepackage{floatflt}
\usepackage{slashed}
\newcommand\be{\begin{equation}}
\newcommand\ee{\end{equation}}

\allowdisplaybreaks[4]

\begin{document}

\title{The Necessity of Multi-Band Observations of the Stochastic Gravitational Wave Background}
\author{S.D. Odintsov$^{1,2}$}\email{odintsov@ice.csic.es}
\author{V.K. Oikonomou,$^{3,4}$}\email{voikonomou@gapps.auth.gr;v.k.oikonomou1979@gmail.com}
\affiliation{$^{1)}$ Institute of Space Sciences (ICE, CSIC) C. Can Magrans s/n, 08193 Barcelona, Spain \\
$^{2)}$ ICREA, Passeig Luis Companys, 23, 08010 Barcelona, Spain\\
$^{3)}$Department of Physics, Aristotle University of
Thessaloniki, Thessaloniki 54124, Greece \\ $^{4)}$L.N. Gumilyov
Eurasian National University - Astana, 010008, Kazakhstan}
%$^{2)}$ Laboratory for Theoretical Cosmology, International Center
%of Gravity and Cosmos, Tomsk State University of Control Systems
%and Radioelectronics  (TUSUR), 634050 Tomsk, Russia

 \tolerance=5000

\begin{abstract}
In this work we highlight an important perspective for the
complete understanding of the stochastic gravitational background
structure. The stochastic gravitational wave background is perhaps
the most important current and future tool towards pinpointing the
early Universe phenomenology related with the inflationary era and
the subsequent reheating era. Many mysteries are inherent to the
stochastic spectrum so in this work we highlight the fact that the
complete understanding of early Universe physics and of
astrophysical processes requires data from many distinct frequency
band ranges. The combination of these data will provide a deeper
and better understanding of the physics that forms the stochastic
gravitational wave background, in both cases that it is of
cosmological or astrophysical origin. We also discuss how the
reheating temperature may be determined by combining multi-band
frequency data from gravitational wave experiments and we also
discuss how the shape of the gravitational wave energy spectrum
can help us better understand the physical processes that formed
it.
\end{abstract}

%PACS numbers: 04.50.Kd, 95.36.+x, 98.80.-k, 98.80.Cq

\maketitle

\section{Introduction}

Undoubtedly, in the post-Higgs detection era, the only realistic
way to study the early Universe is via the stochastic
gravitational wave spectrum of the primordial Universe. Indeed,
particle accelerators have no reports on particles being detected
after the 2012 detection of the Higgs particle
\cite{ATLAS:2012yve}, even for nearly two orders of magnitude
beyond the Higgs mass. The Large Hadron Collider has currently
reached nearly 15$\,$TeV center-of-mass energies and no sign of
new physics in terms of some particle has emerged. Apparently, the
only way to study in a realistic way high energy physics is only
via the stochastic gravitational wave background. Indeed, in the
stochastic gravitational wave spectrum, invaluable information is
inherent to its structure and is expected to probe in a unique way
the early Universe, in a way that it is inaccessible to particle
colliders, at least for the next 80 years.

Thus, the future of high energy physics phenomenology relies
heavily on sky-based observations. Indeed, the stage 4 Cosmic
Microwave Background (CMB) experiments
\cite{CMB-S4:2016ple,SimonsObservatory:2019qwx} expected to
commence in 2027, and also the current (NANOGrav and PTA)
\cite{nanograv,Antoniadis:2023ott,Reardon:2023gzh,Xu:2023wog} and
future gravitational wave experiments
\cite{Hild:2010id,Baker:2019nia,Smith:2019wny,Crowder:2005nr,Smith:2016jqs,Seto:2001qf,Kawamura:2020pcg,Bull:2018lat,LISACosmologyWorkingGroup:2022jok},
like LISA, Einstein Telescope, BBO, DECIGO and so on, are expected
to provide important information about the existence and structure
of the primordial gravitational wave spectrum. More importantly,
the inflationary era will be probed directly or indirectly in a
concrete way by the aforementioned experiments. Indeed, the stage
4 CMB experiments will directly probe the existence of the
inflationary B-modes directly in the CMB polarization, and the
gravitational wave experiments will probe indirectly the
inflationary era by confirming the existence of a stochastic
gravitational wave background which might be due to the
inflationary era.

In 2023, the NANOGrav and the PTA collaborations
\cite{nanograv,Antoniadis:2023ott,Reardon:2023gzh,Xu:2023wog}
confirmed the existence of a stochastic gravitational wave
background, but its existence is by far difficult to be attributed
to some specific mechanism. Two candidate sources exist that may
explain the NANOGrav signal, firstly the astrophysical mergers of
galactic black holes and secondly the signal can be attributed to
a cosmological origin. Although the astrophysical explanation has
several shortcomings, experimental and theoretical, like for
example the lack of a theoretical explanation of the last parsec
problem, for the moment it is too early to provide a well-founded
explanation of the NANOGrav-PTA 2023 signal observation.
Nevertheless, many works highlight or discuss the cosmological
perspective of the 2023 stochastic gravitational wave signal see
for example
\cite{sunnynew,Oikonomou:2023qfz,Cai:2023dls,Han:2023olf,Guo:2023hyp,Yang:2023aak,Addazi:2023jvg,Li:2023bxy,Niu:2023bsr,Yang:2023qlf,Datta:2023vbs,Du:2023qvj,Yi:2023mbm,You:2023rmn,Wang:2023div,Figueroa:2023zhu,Choudhury:2023kam,HosseiniMansoori:2023mqh,Ge:2023rce,Bian:2023dnv,Kawasaki:2023rfx,Yi:2023tdk,An:2023jxf,Zhang:2023nrs,DiBari:2023upq,Jiang:2023qbm,Bhattacharya:2023ysp,Choudhury:2023hfm,Bringmann:2023opz,Choudhury:2023hvf,Choudhury:2023kdb,Huang:2023chx,Jiang:2023gfe,Zhu:2023lbf,Ben-Dayan:2023lwd,Franciolini:2023pbf,Ellis:2023oxs,Liu:2023ymk,Liu:2023pau,Madge:2023dxc,Huang:2023zvs,Fu:2023aab,Maji:2023fhv,Gangopadhyay:2023qjr,Wang:2023sij,Wang:2023ost},
and also
\cite{Schwaller:2015tja,Ratzinger:2020koh,Ashoorioon:2022raz,Choudhury:2023vuj,Choudhury:2023jlt,Choudhury:2023rks,Bian:2022qbh,Kuroyanagi:2020sfw,Maity:2024odg,Vagnozzi:2020gtf,Benetti:2021uea,Haque:2021dha}
and furthermore
\cite{Guo:2023hyp,Yang:2023aak,Machado:2018nqk,Regimbau:2022mdu}.

In this line of research, in this perspective letter we highlight
the fact that the complete understanding of early Universe physics
requires the multi-band study of the stochastic gravitational wave
background. This multi-band study will provide insights for many
physical scenarios which we currently ignore. We shall discuss
these issues in a concrete and formal way and we outline these
here. Firstly, the multi-band detection will determine whether the
signal is of astrophysical origin or of cosmological origin. The
physics of galactic black hole mergers should be essentially the
same across many frequencies ranges, so even at larger frequencies
the same physics should apply. If the stochastic signal is
astrophysical, it should be present across many frequency ranges,
and this depends on the mass of the supermassive black holes. This
existence of the stochastic gravitational wave background across a
large frequency range is not necessarily true for the cosmological
explanation of the stochastic signal. Indeed, the stochastic
signal may be absent in specific frequencies, and this could be of
valuable importance. Regarding the cosmological perspective, the
stochastic gravitational wave background could help us to have
insights about the reheating temperature in the Universe. Also
multi-band detections of gravitational waves may provide us with
insights regarding the polarizations of gravitational waves and
also determine any exotic polarizations. In this letter we
formally discuss these issues in a concrete way.

\section{The Astrophysical Perspective of a Stochastic Gravitational Wave Background}

The detection of the stochastic gravitational wave signal in June
2023, initiated a large stream of studies aiming describing such a
signal. A large portion of the studies is focusing on the
astrophysical explanation of the signal. The NANOGrav and PTA's
focus in frequencies of the nanohertz, but LISA and the other
future gravitational wave experiments will probe frequencies in
the range $10^{-4}-100\,$Hz with the lower frequency probed by
LISA and the higher from DECIGO. Apparently, if the mergers of
supermassive black holes are responsible for the 2023 signal, this
signal should be detected in other frequency ranges, probed by
LISA and the Einstein Telescope, since smaller supermassive black
holes should merge and produce stochastic signals in lower
frequency ranges. Apparently, the absence of such a stochastic
signal is some frequency ranges could be a strong indicator that
the astrophysical explanation of the stochastic signal is not
correct. So if astrophysics is behind the stochastic gravitational
wave signal, the stochastic signal should be present in all
related frequency ranges that probe physics of supermassive black
hole mergers.

Also, a vital ingredient of the astrophysical explanation is the
detection of single supermassive black hole mergers. This
detection should also occur in all related frequency ranges.
Currently no such detection has ever been found in the nanohertz
range, thus this casts doubt on the astrophysical perspective.

These two features are the most important issues related with the
astrophysical perspective. But the astrophysical perspective has
also other issues that is needed to overcome. For example, the
spectral slope of the 2023 NANOGrav signal is approximately
3$\sigma$ off the astrophysical prediction coming from
supermassive black holes mergers \cite{sunnynew}, see also
\cite{Bringmann:2023opz}. Also, the complete absence of
anisotropies \cite{NANOGrav:2023gor} in the 2023 signal, the
incomplete solution to the final parsec problem
\cite{Sampson:2015ada}, makes the astrophysical perspective less
likely compared to the cosmological perspective, based on an
Occam's razor approach. In addition to these problems, it is found
statistically, that cosmological models provide a better fit to
the NANOGrav 2023 stochastic gravitational wave background signal,
than the astrophysical perspective, in Bayes factors to a range
from 10 to 100 \cite{NANOGrav:2023hvm}. However, it is rather too
early to make conclusions on the source of the stochastic
gravitational wave background. We list here the most important
features that will favor an astrophysical explanation of the
stochastic gravitational wave signal.

\begin{itemize}
    \item Presence of a stochastic signal in NANOGrav, LISA,
    and the Einstein Telescope.
    \item Detection of single supermassive black hole mergers in
    all detectors, in NANOGrav, LISA and Einstein Telescope
    frequencies.
    \item Detection of large anisotropies in the stochastic
    gravitational wave signal.
\end{itemize}

\section{The Cosmological Perspectives, Inflation, the Reheating Temperature and the Stochastic Gravitational Wave Background}

Currently, the cosmological explanation of the 2023 nanohertz
stochastic gravitational wave signal seems more plausible,
however, it is by far not certain that this is true and even far
uncertain which cosmological scenario is favorably the mechanism
behind the stochastic gravitational wave signal. In this section
we shall discuss the cosmological perspective of the stochastic
gravitational wave signal.

Many things can be told about the spectrum of the stochastic
gravitational waves and only a multi-band analysis can reveal the
true structure of the spectrum. Many possible scenarios can be
possible, for example, detection in all frequencies, from
nanohertz up to 100$\,$Hz, or absence of the stochastic background
in some frequency ranges, for example in LISA, and presence in
nanohertz and in the Einstein Telescope and so on. Each of these
scenarios has its inherent physics and useful information about
the early Universe can be obtained from the multi-band analysis of
the stochastic gravitational wave background. We shall use an
appropriate example in this section in order to exemplify our
thinking and support our arguments. Prior to that, let us mention
a few things about the inflationary perspective in light of the
NANOGrav observations in 2023. The standard inflationary scenario
with a red-tilted tensor spectrum fails by far to explain the
NANOGrav signal. Also in order for the 2023 detection to be
explained one needs a strongly blue-tilted tensor spectrum with
tensor spectral index over unity and a low reheating temperature
of the order $\mathcal{O}(1-40)\,$GeV
\cite{sunnynew,Oikonomou:2023qfz}. In the context of inflationary
theories, Einstein-Gauss-Bonnet theories can yield a blue-tilted
tensor spectrum, but not such a large tensor spectral index
\cite{Oikonomou:2023qfz}. Only some non-local versions of the
Starobinsky model can yield such a large tensor spectral index.
Thus standard and even blue-tilted inflationary theories by
themselves do not suffice for explaining the NANOGrav signal. One
needs the combination of an abnormal reheating era, with a
broken-power-law energy spectrum, a blue-tilted tensor spectral
index with values beyond unity and a low reheating temperature in
order to explain the NANOGrav signal. But the story does not stop
at the NANOGrav signal, the plot will thicken with future
gravitational waves experiments. Indeed, the NANOGrav signal can
be explained by constrained inflationary theories combined with an
abnormal reheating era, but a plethora of scenarios can explain
cosmologically the NANOGrav signal, like for example cosmic
strings, phase transitions and so on, see for example
\cite{sunnynew,Oikonomou:2023qfz,Cai:2023dls,Han:2023olf,Guo:2023hyp,Yang:2023aak,Addazi:2023jvg,Li:2023bxy,Niu:2023bsr,Yang:2023qlf,Datta:2023vbs,Du:2023qvj,Yi:2023mbm,You:2023rmn,Wang:2023div,Figueroa:2023zhu,Choudhury:2023kam,HosseiniMansoori:2023mqh,Ge:2023rce,Bian:2023dnv,Kawasaki:2023rfx,Yi:2023tdk,An:2023jxf,Zhang:2023nrs,DiBari:2023upq,Jiang:2023qbm,Bhattacharya:2023ysp,Choudhury:2023hfm,Bringmann:2023opz,Choudhury:2023hvf,Choudhury:2023kdb,Huang:2023chx,Jiang:2023gfe,Zhu:2023lbf,Ben-Dayan:2023lwd,Franciolini:2023pbf,Ellis:2023oxs,Liu:2023ymk,Liu:2023pau,Madge:2023dxc,Huang:2023zvs,Fu:2023aab,Maji:2023fhv,Gangopadhyay:2023qjr,Schwaller:2015tja,Ratzinger:2020koh,Ashoorioon:2022raz,Choudhury:2023vuj,Choudhury:2023jlt,Choudhury:2023rks,Bian:2022qbh,Guo:2023hyp,Yang:2023aak,Machado:2018nqk,Regimbau:2022mdu}.
It is vital thus to have the entire multi-band structure of the
stochastic gravitational wave background in order to pinpoint the
correct physical theory that may produce the detected pattern, or
exclude theories from the viable candidate theories. In the future
detected pattern of the stochastic gravitational wave background,
were data from all the detectors will be included, from NANOGrav
to LISA and Einstein Telescope, the shape of the stochastic signal
will be of importance, for example if it has a peak structure or
if it is flat, and also important information regarding the
reheating temperature can be obtained, especially in the case that
the stochastic background has a peak structure. A completely flat
but detectable gravitational wave energy spectrum can be obtained
for example by a standard inflationary $f(R)$ gravity, with an
abnormal reheating era realized again by an $f(R)$ gravity, see
Ref. \cite{Oikonomou:2023qfz} for details. Now a gravitational
wave energy spectrum with a peak can be realized by various
scenarios, and it can be detectable by some experiments and remain
undetectable by other experiments. In such a case, information
about the reheating temperature may be obtained. Let us exemplify
this argument by using an Einstein-Gauss-Bonnet theory with a
blue-tilted tensor spectral index. We shall consider a GW170817
compatible Einstein-Gauss-Bonnet theory in which the gravitational
wave speed is exactly equal to unity in which case the
Gauss-Bonnet coupling function $\xi(\phi)$ satisfies the
constraint $\ddot{\xi}=H\dot{\xi}$. This class of theories was
developed in Refs.
\cite{Oikonomou:2022xoq,Oikonomou:2021kql,Odintsov:2020sqy}, see
also the review \cite{review1}, and the gravitational action is
assumed to be,
\begin{equation}
\label{action} \centering
S=\int{d^4x\sqrt{-g}\left(\frac{R}{2\kappa^2}-\frac{1}{2}\partial_{\mu}\phi\partial^{\mu}\phi-V(\phi)-\frac{1}{2}\xi(\phi)\mathcal{G}\right)}\,
,
\end{equation}
where $R$ denotes the Ricci scalar, $\kappa=\frac{1}{M_p}$ and
$M_p$ is the reduced Planck mass and furthermore $\mathcal{G}$
stands for the Gauss-Bonnet invariant which is
$\mathcal{G}=R^2-4R_{\alpha\beta}R^{\alpha\beta}+R_{\alpha\beta\gamma\delta}R^{\alpha\beta\gamma\delta}$
where $R_{\alpha\beta}$ and $R_{\alpha\beta\gamma\delta}$ denote
the Ricci and Riemann tensors.
\begin{figure}[h!]
\centering
\includegraphics[width=40pc]{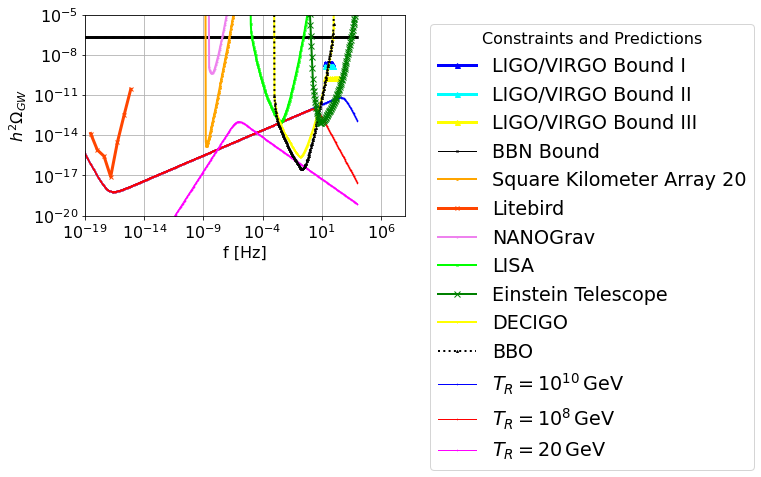}
\caption{The $h^2$-scaled gravitational wave energy spectrum for
the Einstein-Gauss-Bonnet theory at hand versus the various
sensitivity curves of the future gravitational waves experiments,
like the LISA, Einstein Telescope, BBO, DECIGO, for three
reheating temperatures, namely for a high reheating temperature
$T_R=10^{10}\,$GeV, an intermediate reheating temperature
$T_R=10^{8}\,$GeV and a relatively low reheating temperature
$T_R=20\,$GeV, for $n_{\mathcal{T}}\sim 0.3790$.}\label{plot2}
\end{figure}
According to Refs.
\cite{Oikonomou:2022xoq,Oikonomou:2021kql,Odintsov:2020sqy}, if
the gravitational wave speed is equal to unity in natural units,
the slow-roll indices of the inflationary theory acquire the form
\cite{Oikonomou:2021kql},
\begin{align}
&\epsilon_1\simeq\frac{\kappa^2
}{2}\left(\frac{\xi'}{\xi''}\right)^2,\, \, \,
\epsilon_2\simeq1-\epsilon_1-\frac{\xi'\xi'''}{\xi''^2},\,\,\,
\epsilon_3=0,\,\,\,
\epsilon_4\simeq\frac{\xi'}{2\xi''}\frac{\mathcal{E}'}{\mathcal{E}}\,,\\
\notag & \epsilon_5\simeq-\frac{\epsilon_1}{\lambda},\, \,\,
\epsilon_5(1-\epsilon_1)\, ,
\end{align}
where $\mathcal{E}=\mathcal{E}(\phi)$ and $\lambda=\lambda(\phi)$
and,
\begin{equation}\label{functionE}
\mathcal{E}(\phi)=\frac{1}{\kappa^2}\left(
1+72\frac{\epsilon_1^2}{\lambda^2} \right),\,\, \,
\lambda(\phi)=\frac{3}{4\xi''\kappa^2 V}\, .
\end{equation}
Accordingly, the inflationary observational indices are,
\begin{equation}
\label{spectralindex} \centering
n_{\mathcal{S}}=1-4\epsilon_1-2\epsilon_2-2\epsilon_4\, ,
\end{equation}
\begin{equation}\label{tensorspectralindex}
n_{\mathcal{T}}=-2\left( \epsilon_1+\epsilon_6 \right)\, ,
\end{equation}
\begin{equation}\label{tensortoscalar}
r=16\left|\left(\frac{\kappa^2Q_e}{4H}-\epsilon_1\right)\frac{2c_A^3}{2+\kappa^2Q_b}\right|\,
,
\end{equation}
and the sound speed $c_A$ is,
\begin{equation}
\label{sound} \centering c_A^2=1+\frac{Q_aQ_e}{3Q_a^2+
\dot\phi^2(\frac{2}{\kappa^2}+Q_b)}\, ,
\end{equation}
and furthermore,
\begin{align}\label{qis}
& Q_a=-4 \dot\xi H^2,\,\,\,Q_b=-8 \dot\xi H,\,\,\,
Q_t=F+\frac{Q_b}{2},\\
\notag &  Q_c=0,\,\,\,Q_e=-16 \dot{\xi}\dot{H}\, .
\end{align}
Hence, the tensor-to-scalar ratio and the tensor spectral index
acquire the form,
\begin{equation}\label{tensortoscalarratiofinal}
r\simeq 16\epsilon_1\, ,
\end{equation}
\begin{equation}\label{tensorspectralindexfinal}
n_{\mathcal{T}}\simeq -2\epsilon_1\left ( 1-\frac{1}{\lambda
}+\frac{\epsilon_1}{\lambda}\right)\, .
\end{equation}
A viable model has the following Gauss-Bonnet coupling function
\cite{Oikonomou:2021kql},
\begin{equation}
\label{modelA} \xi(\phi)=\beta  \exp \left(\left(\frac{\phi
}{M}\right)^2\right)\, ,
\end{equation}
with $\beta$ being a dimensionless parameter, and also $M$ stands
for a free parameter with mass dimensions $[m]^1$. The
corresponding scalar potential is constrained to have the
following form during the slow-roll era,
\begin{equation}
\label{potA} \centering V(\phi)=\frac{3}{3 \gamma  \kappa ^4+4
\beta  \kappa ^4 e^{\frac{\phi ^2}{M^2}}} \, ,
\end{equation}
with $\gamma$ being a dimensionless integration constant.
Accordingly, the slow-roll indices are found to be,
\begin{align}
& \epsilon_1\simeq \frac{\kappa ^2 M^4 \phi ^2}{2 \left(M^2+2 \phi
^2\right)^2}, \, \,\, \epsilon_2\simeq \frac{M^4 \left(2-\kappa ^2
\phi ^2\right)-4 M^2 \phi ^2}{2 \left(M^2+2 \phi ^2\right)^2},\,
\,\,\epsilon_3=0\,,
\end{align}
\begin{align}
\epsilon_5\simeq -\frac{4 \beta  \phi ^2 e^{\frac{\phi
^2}{M^2}}}{\left(M^2+2 \phi ^2\right) \left(3 \gamma +4 \beta
e^{\frac{\phi ^2}{M^2}}\right)} ,\, \,\, \epsilon_6\simeq -\frac{2
\beta  \phi ^2 e^{\frac{\phi ^2}{M^2}} \left(M^4 \left(2-\kappa ^2
\phi ^2\right)+8 M^2 \phi ^2+8 \phi ^4\right)}{\left(M^2+2 \phi
^2\right)^3 \left(3 \gamma +4 \beta  e^{\frac{\phi
^2}{M^2}}\right)} \, ,
\end{align}
and the corresponding inflationary take the form,
\begin{align}\label{spectralpowerlawmodel}
& n_{\mathcal{S}}\simeq -1-\frac{\kappa ^2 M^4 \phi
^2}{\left(M^2+2 \phi ^2\right)^2}+\frac{4 \phi ^2 \left(3 M^2+2
\phi ^2\right)}{\left(M^2+2 \phi ^2\right)^2}\\ & \notag
+\frac{4608 \beta ^2 \phi ^6 e^{\frac{2 \phi ^2}{M^2}} \left(6
\gamma  \phi ^2+16 \beta  e^{\frac{\phi ^2}{M^2}} \left(M^2+\phi
^2\right)+9 \gamma  M^2\right)}{\left(M^2+2 \phi ^2\right)^4
\left(3 \gamma +4 \beta  e^{\frac{\phi ^2}{M^2}}\right)^3} \, ,
\end{align}
\begin{align}\label{tensorspectralindexpowerlawmodel}
& n_{\mathcal{T}}\simeq \frac{\phi ^2 \left(-4 \beta e^{\frac{\phi
^2}{M^2}} \left(M^4 \left(3 \kappa ^2 \phi ^2-2\right)+\kappa ^2
M^6-8 M^2 \phi ^2-8 \phi ^4\right)-3 \gamma \kappa ^2 M^4
\left(M^2+2 \phi ^2\right)\right)}{\left(M^2+2 \phi ^2\right)^3
\left(3 \gamma +4 \beta  e^{\frac{\phi ^2}{M^2}}\right)}
 \, ,
\end{align}
\begin{equation}\label{tensortoscalarfinalmodelpowerlaw}
r\simeq \frac{8 \kappa ^2 M^4 \phi ^2}{\left(M^2+2 \phi
^2\right)^2}\, .
\end{equation}
So the model is viable and has a blue-tilted tensor spectral index
in the range $n_{\mathcal{T}}=[0.378856,0.379088]$ and a
tensor-to-scalar ratio $r\sim 0.003$, for the following values of
the free parameters $\mu=[22.09147657871,22.09147657877]$,
$\beta=-1.5$, $\gamma=2$, for $N=60$ $e$-foldings. The energy
spectrum of the primordial gravitational waves has the following
form \cite{Odintsov:2021kup},
\begin{align}
\label{GWspecfR}
    &\Omega_{\rm gw}(f)=\frac{k^2}{12H_0^2}r\mathcal{P}_{\zeta}(k_{ref})\left(\frac{k}{k_{ref}}
\right)^{n_T} \left ( \frac{\Omega_m}{\Omega_\Lambda} \right )^2
    \left ( \frac{g_*(T_{\rm in})}{g_{*0}} \right )
    \left ( \frac{g_{*s0}}{g_{*s}(T_{\rm in})} \right )^{4/3} \nonumber  \left (\overline{ \frac{3j_1(k\tau_0)}{k\tau_0} } \right )^2
    T_1^2\left ( x_{\rm eq} \right )
    T_2^2\left ( x_R \right )\, ,
\end{align}
so in Fig. \ref{plot2} we plot the $h^2-$scaled energy spectrum of
the blue-tilted Einstein-Gauss-Bonnet theory for three reheating
temperatures, namely for a high reheating temperature
$T_R=10^{10}\,$GeV, an intermediate reheating temperature
$T_R=10^{8}\,$GeV and a relatively low reheating temperature
$T_R=20\,$GeV, for $n_{\mathcal{T}}\sim 0.3790$. Now the physical
picture in the plot of Fig. \ref{plot2} is very clear. The high
and intermediate reheating cases can be marginally detected by the
LISA experiment, while can be both detectable by the BBO, DECIGO
and the Einstein Telescope experiments. The low reheating
temperature scenario though is detectable only by the BBO and thus
remains undetectable by all the rest interferometers. This
physical picture is exactly the kind of example we wanted to
highlight, since if such a pattern is detected, this can clearly
give us hints about the reheating temperature. If for example no
detection of stochastic signal occurs in the LISA and Einstein
Telescope, with a signal detected though by BBO or even marginally
by DECIGO, clearly this can point out that an inflationary
background may be responsible for the stochastic gravitational
wave background, but with a very low reheating temperature. This
also clearly shows the necessity of a multi-band frequency
analysis of the stochastic gravitational wave background. Thus,
currently we are by far back towards understanding what is the
source of the 2023 NANOGrav detection and the necessity of more
gravitational wave experiments is clearly highlighted. In the case
that the signal is detected by all interferometers with the same
amplitude, this would clearly indicate a flat energy spectrum thus
an inflationary modified gravity with an abnormal reheating era
might be the source of the stochastic gravitational wave
background. Also the implications of a low reheating temperature
would put in peril the electroweak phase transition, since
reheating temperatures of at least 150$\,$GeV are needed. Thus in
this case many new problems would emerge. Clearly in this case,
the plot heavily thickens. We list here the most important
arguments of the cosmological perspective:

\begin{itemize}
    \item The importance of the shape of the spectrum, is it flat or it is consisted by
    peaks.How many peaks, how many experiments detect the signal
    and what is the amplitude in each detection.
    \item If some interferometers detect the signal while others
    do not, can we make the conclusion that a low-reheating
    temperature was achieved in the Universe?
    \item In the case of low reheating, new problems arise
    regarding the electroweak phase transition. Perhaps a new
    mechanism for the electroweak phase transition is needed?
    \item Depending on the amplitude, does the signal imply an
    abnormal reheating era with a broken-power-law form in the
    frequency range probed by the future gravitational wave
    experiments?
    \item Which cosmological explanation answers the most of the
    above questions in an optimal way? Choosing the most favorable
    theory, or combination of theories is the ultimate quest for
    current and future theoretical physicists.
\end{itemize}

\section{Exotic Polarizations and Ordinary Polarizations of the Gravitational Waves}

Of great importance is the exact determination of the
gravitational wave polarizations, especially the exotic
polarizations predicted by modified gravity theories, such as the
scalar polarizations of $f(R)$ gravity, which are completely
absent in ordinary General Relativity. Currently in LIGO-Virgo
detectors, the technology already used does not allow to even
distinguish between the two standard tensor polarization, thus
even asking the question for extra polarization is unrealistic and
by far a formidable task. This is not the case though for future
gravitational wave experiments. An important feature is to use
this future multi-band detectors in order to find extra
gravitational wave polarizations. This is a formidable task
especially for detectors functioning in different frequency
ranges, when we are discussing about single astrophysical black
hole mergers. This can work in favor us in the following way, if
for example a detection is made in the LISA frequencies, one can
expect the detection to appear in the higher frequency detector
like the Einstein Telescope at a later time, and the combination
of the data of the two detectors may reveal the presence of extra
polarizations, like scalar polarizations, and even distinguish
among tensor polarizations, or even reveal circular polarizations.
Thus once more the necessity of using multi-band gravitational
wave analysis is highlighted.

\section*{Concluding Remarks and Discussion}

In this letter we explained why it is of fundamental importance to
have data on the gravitational waves from many gravitational wave
detectors that function in distinct frequency range. As we
explained, this is very important since it will allow us to reveal
physics of the early Universe that cannot be revealed from
terrestrial particle colliders. Also it will allow us to better
understand the propagating degrees of freedom of the gravitational
waves. As we pointed out, in single astrophysical mergers,
multi-band analysis can reveal extra and exotic polarizations of
the gravitational waves, such as scalar or circular polarizations,
or even distinguish among the existing tensor polarizations. Also
regarding the astrophysical perspective of the stochastic
gravitational wave background, a multi-band analysis will allow us
to understand that the supermassive black holes mergers actually
occur. Important information regarding the reheating temperature
can be revealed by the presence or absence of a signal from future
interferometers and also the shape of the overall spectrum can
show us which theories may describe the background mechanism that
generates the signal. The detection of a flat signal with the same
amplitude for example can indicate some red-tilted modified
gravity theory with an abnormal reheating era and so on. As it
seems, the future is fruitful for gravitational wave physics and
more gravitational wave detectors are needed in order to solve the
major astrophysical and cosmological puzzles. Also it is worth
having in mind future experiments not considered here and their
perspective in detecting gravitational wave signals
\cite{Campeti:2020xwn}. Furthermore, the possibility of testing
inflation and the mysterious reheating era with GHz gravitational
waves is also something worth having in mind for future research
directives, see for example
\cite{Ringwald:2020ist,Vagnozzi:2022qmc,Bernal:2023wus,Ghiglieri:2022rfp,Barman:2023rpg,Barman:2023ymn,Klose:2022knn,Garcia:2024zir}.

\section*{Acknowledgments}

This work was partially supported by MICINN (Spain), project
PID2019-104397GB-I00, and by the program Unidad de Excelencia
Maria de Maeztu CEX2020-001058-M, Spain (S.D.O). This research has
been is funded by the Committee of Science of the Ministry of
Education and Science of the Republic of Kazakhstan (V.K.O) (Grant
No. AP19674478).

\end{document}